\documentclass[usenatbib]{mn2e}
 \usepackage{color}
 \usepackage{float} 
 \usepackage[fleqn]{amsmath}
 \usepackage{epsfig,floatflt}
 \usepackage{natbib}
 \usepackage{subfigure}
 \usepackage{amssymb}
 \usepackage{multirow}
 \usepackage{ulem}
 \usepackage{bm}
 \usepackage{url}
 \usepackage{placeins}

\newcommand{\aap}{A\&A}

\newcommand{\apj}{ApJ}
\newcommand{\apjs}{ApJS}

\newcommand{\mnras}{MNRAS}
\newcommand{\araa}{ARA\&A}
\newcommand{\apjl}{ApJL}

\title{What size halos do local LIRGs live in?}

\author[Abiy G. Tekola et al.]
{Abiy G. Tekola$^{1,2,3}$, Andreas A. Berlind$^4$, Petri V\"{a}is\"{a}nen$^2$ \\
$^1$Las Cumbres Observatory Global Telescope Network, Goleta, CA, 93117, USA\\
$^2$South African Astronomical Observatory, PO Box 9, 7935 Observatory, Cape Town, South Africa\\
$^3$Astronomy Department, and Astrophysics, Cosmology and Gravity Centre (ACGC), University of Cape Town,\\ Private Bag X3, Rondebosch 7701, South Africa\\
$^4$Department of Physics and Astronomy, Vanderbilt University, Nashville, TN 37235, USA}

\begin{document}
\maketitle
\begin{abstract}
This work investigates the preferred environment of local Luminous IR Galaxies (LIRGs) in terms of the host halos that they inhabit, and in comparison to a control galaxy sample.  The LIRGs are drawn from the IRAS Point Source Catalogue redshift survey (PSCz), while the control sample is drawn from the 2MASS redshift survey (2MRS).  A friends-of-friends algorithm was run on the 2MRS sample to identify galaxies living in the same dark matter halos and the PSCz galaxies were then associated with these identified halos.  We show that the relative probability of finding local LIRGs with respect to 2MASS galaxies is largest in approximately group size halos ($M_{halo} \sim 10^{13}M_\odot$), and declines both in the cluster regime and in smaller halos.  This confirms, using a different technique than in previous work, that local LIRGs are indeed more abundant in group environments than elsewhere.  We also find a trend between the $L_{IR}$ values of LIRGs and their location within their host dark matter halos, such that the average location of LIRGs with high IR luminosity is closer to the halo centre than for low IR luminosity galaxies.  Moreover, this trend does not seem to depend on halo mass.\\

\end{abstract}

\section{\large Introduction}
The extreme cases of star formation (SF) in the Universe are exhibited by classes of galaxies such as Luminous Infrared Galaxies (LIRGs; defined as galaxies with IR luminosities of $10^{11}L_
\odot \leq L_{IR} \leq 10^{12}L_\odot$) and Ultra-Luminous IR Galaxies (ULIRGs; $10^{12}L_\odot \leq L_{IR} \leq 10^{13}L_\odot$) and as a result, it is impossible to discuss the phenomenon of extreme SF in the Universe without the mention of LIRGs and ULIRGs.  These galaxies are good laboratories to attempt to understand the nature and detailed mechanism of these exotic forms of SF and they have been the focus of many studies \citep[see e.g.][and references therein]{sanders1996,elbaz2011}. 

It has long been clear that very strong star forming galaxies in the local Universe are related to mergers and interactions  \citep[e.g.][and references therein]{Tekola2012}. Their IR output correlates with the level of their interaction and merger stage, with low-$L_{IR}$ galaxies being often associated with early stages of interactions while the high-$L_{IR}$ ones, especially ULIRGs, are usually in later stages of merging, or are merger remnants. 

At higher redshifts the situation is less clear, however.  Some observational studies show that high-$z$ LIRGs and ULIRGs are in fact forming stars in more modest and quiescent fashions with minimal contribution from interactions and mergers \citep{bell2005, elbaz2007}.   Herschel Space Observatory studies have revealed that levels of SF which would be considered extreme locally originate in the "Main Sequence" of spiral-like star forming galaxies defined by a tight SFR-($M_\ast$) trend in the high redshift Universe \citep{elbaz2011,Rodighiero2011,Noeske2007, pannella2009, karim2011}.  This SF scenario is also supported by numerical simulations, according to which steady accretion of cold gas is the driving mechanism for high SF rates, and consequently LIRGs and ULIRGs at high redshift should be dominated by extended disky morphology and more quiescent form of SF \citep{dekel09, keres2005}.  On the other hand, there are studies indicating that extreme SF at high-$z$ is not much different from the local Universe \citep{Kartaltepe2011} in that a significant fraction of LIRGs and ULIRGs experience a violent central starburst suggesting the importance of mergers and interactions to acquire their gas. These  studies further revealed that the $L_{IR}$ output of these classes of galaxies at $z\sim 1$ is tightly correlated with their morphology, like their local counterparts: the fraction of mergers and interaction increases with $L_{IR}$ while the fraction of disk-like objects declines with $L_{IR}$.
 
From the discussion above, it is clear that galaxy interaction and merger are some of the mechanisms driving extreme form of star formation though the extent of their influence as a function of cosmic time is arguable. This sends a clear message that galaxy environment has a role to play in the matter. Global environment is one form of environment that has influence on extreme form of star formation \citep[see e.g.][]{Tekola2012} and hence understanding the global environment in detail can shed light on the mechanisms of the extreme form of star formation in the local Universe which, in turn, helps to construct a clear picture of its time evolution. 

The global environment of LIRGs is not generally much investigated to date in stark contrast to studies conducted on their immediate surroundings (within tens of Kpc) and on their one-to-one interaction \citep[][and references therein]{ellison2013}. The few studies conducted so far were based on number counts of galaxies to measure global environments of LIRGs, and they seem to agree that these galaxies do not live in cluster environments.  \cite{Tekola2012} studied the Mpc scale global environment of LIRGs and concluded that they preferentially live in group environments and avoid cluster and field environments.  Another study by \cite{tacconi2002} concluded that local LIRGs and ULIRGs do not live in cluster environments. This work attempts to further constrain the global environments of LIRGs using a different approach. Instead of the traditional galaxy count, we use the sizes of host dark matter halos as a measure of global environments.

\section{Data and Methods}
\subsection{Galaxy samples}

In this work, we have made use of volume-limited samples extracted from two redshift surveys: the IRAS Point Source redshift survey  (PSCz; \citealt{saunder2000}) and the 2MASS redshift survey (2MRS; \citealt{huchra2012}).  Our main objective is to study the local environments of IR galaxies, especially LIRGs, in terms of the size of the dark matter halos that they are associated with.  The IR-selected sample from the IRAS PSCz catalogue is used as the target sample, and is hereafter referred to as the target galaxy sample.  The K-band selected sample from the 2MRS was used to identify groups based on the mass of their dark matter halo, and is hereafter referred to as the density field sample. This 2MRS sample was also used as a control galaxy sample to compare with the target sample.

We constructed volume-limited samples from the 2MRS survey in order to satisfy two requirements. First, the volume enclosed by the volume-limited sample redshift limit must have a maximum number of 2MRS galaxies, and second, when the same redshift limit is applied to the target galaxy sample, the resulting volume-limited PSCz sample must have a statistically reasonable number of LIRGs.  Using these criteria, we selected a 2MRS volume-limited sample with $z \le0.024$ and $M_K\le-23.8$.  In order to probe down to lower halo masses, we also chose a second sample with $z\le0.02$ and $M_K\le-23.4$.  Hereafter, these $z\le0.02$ and 0.024 samples are referred to as V2M1 and V2M2 and they contain 5044 and 5457 galaxies, respectively. 

\begin{figure}
\centering
\includegraphics[scale=0.35]{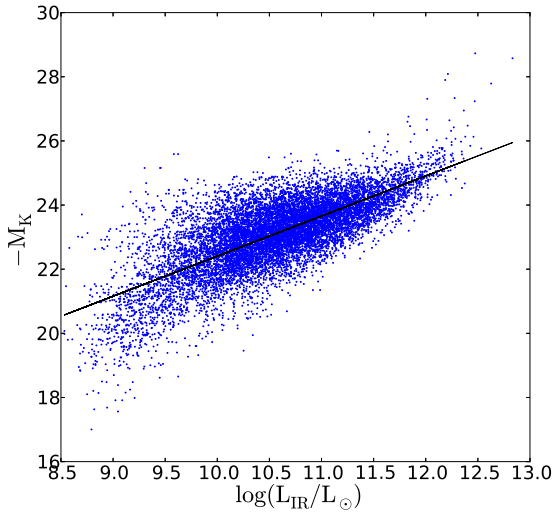}
\caption{The relation between $M_K$ and $L_{IR}$ values of PSCz galaxies. The $M_K$ values are estimated from matching the PSCz galaxies to the 2MASS survey. The line represents the best fit, and is used to decide the luminosity limit of the IRAS galaxiy volume-limited samples.}
\label{fig: mk_lir}
\end{figure}

For both 2MRS volume-limited samples, we constructed corresponding target galaxy volume-limited samples from PSCz. The maximum redshift for each sample was fixed to that of the corresponding density field 2MRS sample described above.  The choice of the minimum $L_{IR}$ limit for these target galaxy samples was made in such a way as to include the LIRGs of interest in the samples, while at the same time ensuring that the PSCz and the 2MRS magnitude limits are roughly similar.  In order to do this, we obtained K-band magnitudes for the PSCz galaxies by matching to the full 2MASS survey, and then studied the resulting relation between $M_K$ and $L_{IR}$ (see Fig.~\ref{fig: mk_lir}).  A simple fit to this data yields the relation $log(L_{IR}/L_\odot)=-0.8 M_K-7.9$.  While there is obvious scatter in this relation due to different populations of galaxies in the sample, it is satisfactory for the purpose of setting a statistical flux limit for the sample; no physical parameters are derived from the relation.  According to this relation, our 2MRS sample limit of $M_K=-23.4$ corresponds to an IR luminosity of $log(L_{IR})=10.8$, which is safely higher than the PSCz completeness limit in these volumes and is close to the nominal LIRG limit of $log(L_{IR}) = 11$. Therefore, $\log(L_{IR}) = 10.8$ was chosen as the luminosity limit for both PSCz volume-limited samples. This ensures that the K-band luminosities of our target and control samples are roughly similar.  We have checked that the exact choice of the $L_{IR}$ limit within the scatter range evident in the Figure~\ref{fig: mk_lir} makes no difference to the final results. The PSCz volume-limited samples are hence characterized by the following redshift and $L_{IR}$ limits: $z\le0.02$ and $log(L_{IR}/L_\odot)\ge10.8$, and $z\le0.024$ and $log(L_{IR}/L_\odot)\ge10.8$.  Hereafter, we refer to these samples as VP1 and VP2 respectively.  The magnitude and redshift limits of all samples described above are summarized in Table~\ref{tab:mk_vol_table1}.

\begin{table}
\centering
\begin{tabular}{c c c c }
\hline\hline
Name & Magnitude limit & $z_{max}$ & No. of galaxies \\[0.5ex]
\hline
V2M1 & $M_K\le-23.4$ & 0.02 & 5044 \\
V2M2 & $M_K\le-23.8$ & 0.024 & 5457 \\
Vp1& $L_{IR} \ge 10.8$ & 0.02 & 290\\
Vp2& $L_{IR} \ge 10.8$ & 0.024 & 565\\
\\
\hline
\end{tabular}
\caption{The volume-limited samples constructed from the 2MRS and PSCz surveys.}
\label{tab:mk_vol_table1}
\end{table}

\subsection{Constructing the group catalogue and assigning halo mass}

Groups from the two 2MRS volume-limited samples were identified using the well-known Friends-of-Friends $(FoF)$ group finding algorithm \citep[see][]{huch1982, geller1983}. The algorithm is specifically custom-designed by \cite{berlind2006} so that it groups together galaxies living in the same dark matter halo.  Table~\ref{tab: identifid_systems} shows the breakdown of the total number of identified systems into isolated galaxies, pairs, and systems with more than two galaxies.
 
\begin{table}
\centering
\begin{tabular}{c c c c c}
\hline\hline
Name & Isolated & pairs & $\ge$ 3 galaxies \\[0.5ex]
& systems &&&\\[0.5ex]
\hline
V2M1 & 2291  & 882 & 1871\\
V2M2 & 2592  & 982 & 1883\\
\\
\hline
\end{tabular}
\caption{The distribution of the galaxy systems identified from the galaxy density sample using the Friends-of-Friends algorithm.}
\label{tab: identifid_systems}
\end{table}

The K-band group luminosity is computed for each of the identified groups by summing up the K-band light of the individual detected galaxies in the groups. The total K-band absolute magnitude of a group is thus defined as 
\begin{equation} 
M_{K_{total}}=-2.5\,log(\sum_i 10^{-0.4\,M_{K_i}}),
\end{equation} 
where $M_{K_i}$ are the K-band absolute magnitudes of the individual galaxies in a group.  This group magnitude is not equal to the actual total magnitude of the group because the underlying 2MRS samples are only complete to a certain absolute magnitude limit.  We thus only use these computed group magnitudes to determine the rank order of groups according to K-band luminosity.  We then assume a monotonic relation between a group's luminosity and the mass of its underlying dark matter halo to assign rough virial masses to our groups \citep[see e.g.][]{quintero2005, berlind2006, Tekola2012}. Specifically, we match the measured space density of ranked groups to the theoretical space density of dark matter halos that is predicted by the concordance cosmological model \citep{Warren2006}. This technique is usually referred to as ``abundance matching''.  We note that this method assumes no scatter in mass at a given luminosity and the result is only meant to produce approximate statistical values of halo mass.
 
\begin{figure*}
\centering
\includegraphics[scale=0.54]{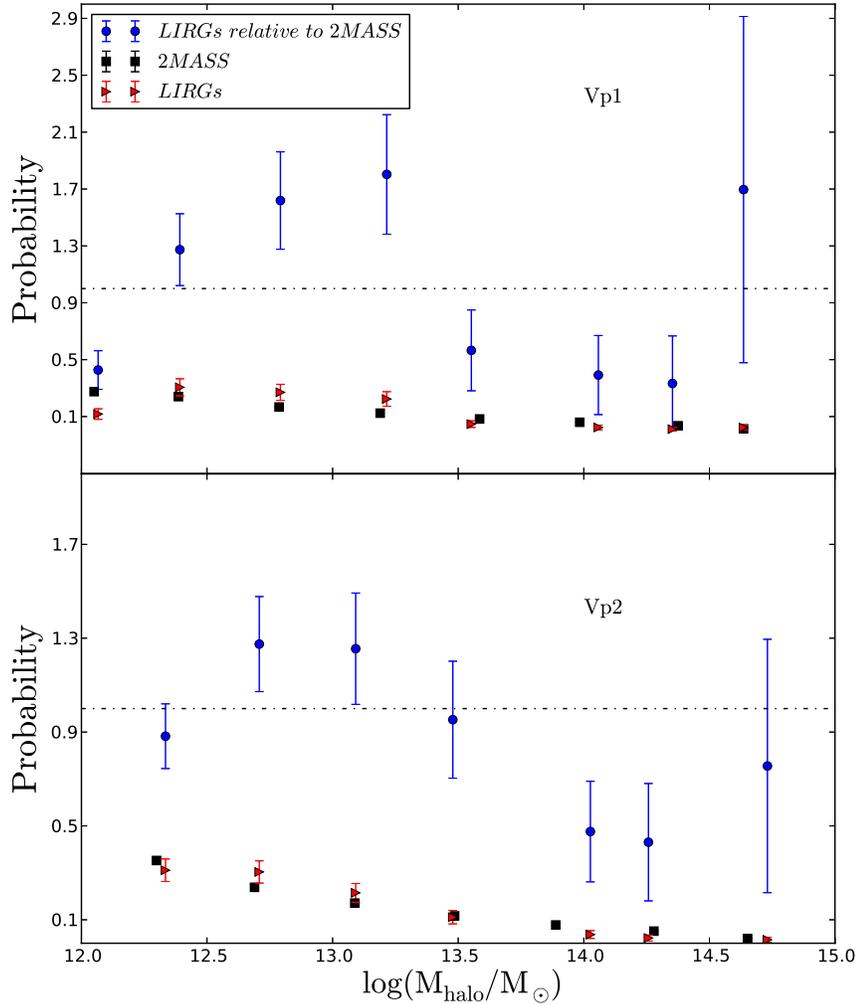}
\caption{The binned probability distribution of dark matter halo mass corresponding to LIRGs (triangles) and 2MASS galaxies (squares), as well as the ratio of these two probabilities (circles).  The two panels show results for the two separate volume-limited samples. Error bars are estimated assuming Poisson statistics.}
\label{fig: 2mrs_groupmember_bin}
\end{figure*}

The Right Ascension and Declination of the groups are defined by the centroid of the group members, and the mean redshift $<z>_g$ of the group is calculated from a simple mean of all the individual members $z_i$. Since groups of galaxies appear stretched in the line-of-sight direction due to radial peculiar velocities of member galaxies, we assume that the identified groups have an approximate cylindrical geometry with a radius equal to their virial radius ($R_{vir}$) given by \citep{gunn1972} and presented as
\begin{equation}
R_{vir}=(\frac{(3/4)M_{vir}}{200\pi\bar{\rho}})^{1/3},
\end{equation}
where $M_{vir}$ is the virial mass and $\bar\rho$ is the mean density of the Universe. The length of the cylinder is two times the maximum rotational velocity of the halo, 2$V_{max}$, where $V_{max}$ is determined from the rotational velocity profile that relates velocity and radius. 

\subsection{Correlating individual PSCz galaxies with identified 2MRS groups}

Since our ultimate goal in this work is to find out what type of environment, or halo size, IRAS PSCz galaxies are preferentially living in, we correlate the individual PSCz galaxies with the 2MRS group catalogue in order to identify the host halo for each target galaxy.  We check whether a particular PSCz galaxy belongs to a particular 2MRS group by calculating the distance of the galaxy from the group centre and comparing to the group size.

Suppose a target IRAS galaxy in question has $d_{\parallel}$ and $d_{\perp}$ as the line-of-sight and perpendicular distances respectively from the geometric centre of a particular group. $d_\parallel$ is then given by
\begin{equation}
d_\parallel=(c/H_o)|z_i-<z>_g|.
\end{equation}
For a small angle $\theta$ between the line of sight to the group and the galaxy in question, $d_\perp$ can be approximated to good accuracy as
\begin{equation}
d_{\perp}\approx(c/H_o)<z>_g\theta. 
\end{equation}
Both distances are given in units of $h^{-1}Mpc$. A galaxy is considered to be part of a certain group if it lies inside the cylinder defined by $d_\parallel$ and $d_\perp$ as its length and radius, respectively. This in other words means that a galaxy belongs to a certain group if $d_\parallel$ $\le$ $V_{max}/H_o$ and $d_\perp$ $\le$ $R_{vir}$. In cases where a galaxy might lie in the cylinder of more than one group, then the three-dimensional distance from each group to the galaxy weighted by the $R_{vir}$ and $V_{max}/H_0$ is calculated as
\begin{equation}
d=\sqrt{(H_0 d_\parallel/V_{max})^2+(d_\perp/R_{vir})^2},
\end{equation}  
and the galaxy finally belongs to the group with the smallest value of $d$.  

We repeat the exact same procedure replacing the PSCz target sample with the 2MRS control sample.  In other words, we associate each 2MRS control galaxy with a group using these cylindrical criteria.  This may seem strange given that 2MRS galaxies were already associated with groups during the group-finding.  However, the cylindrical linking volume that we use for the target sample is not equal to the effective volume resulting from the friends-of-friends algorithm, and it is important to use the same exact criteria for the target and control samples so that they may be directly compared.

 \begin{figure*}
\centering
\includegraphics[scale=0.5]{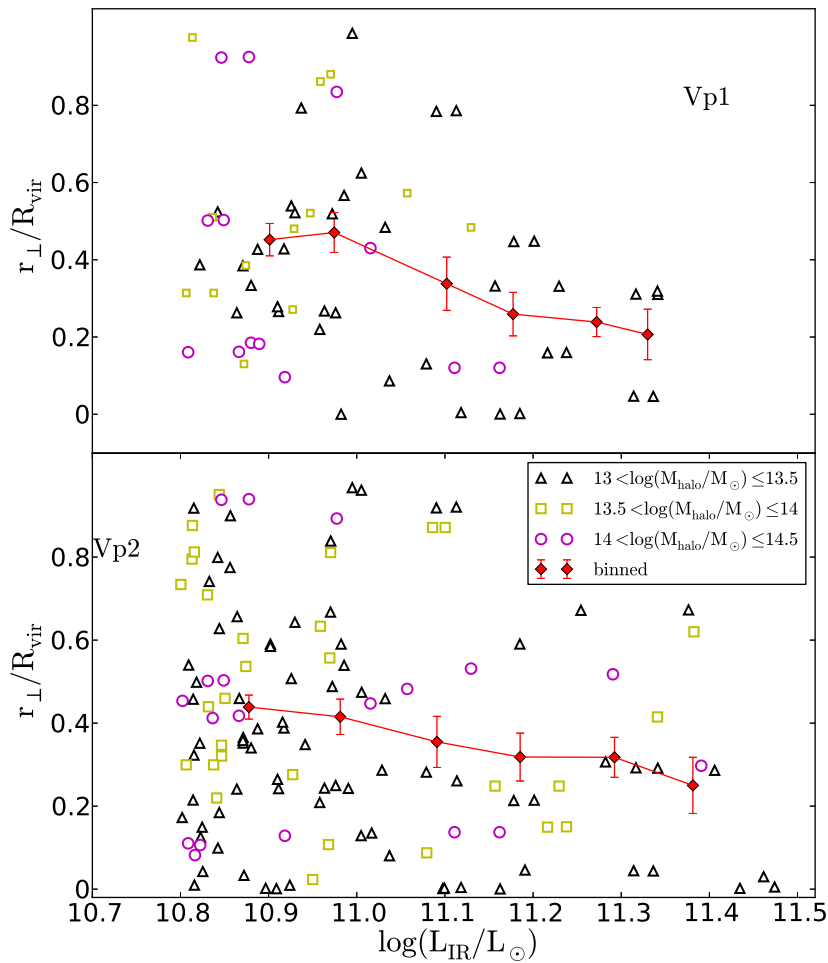}
\caption{The distribution of the projected distances of LIRGs from the galaxy system centre in units of the virialized radius versus their $L_{IR}$. The red lines represent the binned values in $L_{IR}$ and the different markers represent the different halo mass bins. The data in the both panels cover all IRAS galaxies in the volume limited target samples, i.e. they all have $> 10^{10.8}L_\odot$.  The error bars with the averaged points are estimated from Poisson statistics in a bin of $log(L_{IR}/L_\odot)=0.1$ size.}
\label{fig: dis_cen_includingfield}
\end{figure*} 

\section{Results}

\subsection{Distribution of host halo masses}

Now that we have assigned halo masses to both target and control galaxy samples, we can study their halo environments. Note that we do not included cases of isolated IRAS PSCz galaxy systems since no halo mass can be associated to them. This will not change the trends observed in the plotted range in any way, however. We estimate the probability distribution of halo masses for PSCz galaxies by taking the ratio of the number of PSCz galaxies in halo mass bins to the total number of PSCz galaxies. The result is shown in red triangles in the two panels of Figure~\ref{fig: 2mrs_groupmember_bin}. The trends indicate that the probabilities decline with the size of the galaxy systems they are associated with.  At the higher mass end (cluster regime) the probability is much smaller than at the lower mass end.  We also estimate a similar probability distribution for the 2MRS control samples in order to make a comparison of the LIRG's environment with that of the general population of galaxies.  These probability distributions are shown as black squares in the panels of Figure~\ref{fig: 2mrs_groupmember_bin} and they exhibit similar trends as the LIRGs.

The declining probability with halo mass observed both for the LIRGs and the 2MRS samples is simply a reflection of the deficiency of high mass systems (such as clusters) as compared to low mass systems \citep{press1974,bond1991}. In order to eliminate this effect from the result and also to make a comparison between the environments of LIRGs and 2MASS galaxies, we take the ratio of their probability distributions.  The blue solid circles in both panels of Figure~\ref{fig: 2mrs_groupmember_bin} show the relative probability of finding LIRGs in a given halo mass with respect to 2MRS control sample galaxies. Error bars on all points are calculated assuming Poisson statistics.  These results clearly show that LIRGs do not simply trace 2MASS galaxies in their distribution of host halo masses.  Instead, LIRGs have a preference for halos of intermediate (group size) masses, avoiding both smaller (Milky Way size) and larger (cluster size) halos relative to normal 2MASS galaxies.  This effect is fairly large, with the fraction of LIRGs in group size systems being as much as $\sim 70$\% larger than the fraction of 2MASS galaxies in similar size systems.  On the other hand, the fraction of LIRGs in clusters or the field are as much as $\sim 50$\% lower than that for normal galaxies.

The two panels of Figure~\ref{fig: 2mrs_groupmember_bin} show that results for the two volume-limited samples are slightly different despite the fact that the LIRGs in both cases are the same galaxy types. If halo finding and mass assignment were perfect then a halo found using a fainter 2MRS sample would also be found with the brighter sample (with fewer galaxies per halo) and they would result in the same mass. The only difference would be some lower mass halos in the fainter sample not being found in the brighter sample, which is reflected by the increasing lower-limit of halo masses going from the top to the bottom panel in Figure~ \ref{fig: 2mrs_groupmember_bin}.   However, since the halos are found using two different 2MRS samples, the differences seen at higher masses must be caused by halo identification and mass assignment errors, and possibly also by real cosmic variance since the volumes probed are somewhat different.  Nevertheless, the basic trends remain the same in both samples and are thus likely robust against such effects.

\subsection{Distance from the centre of galaxy groups}

In the previous section we discussed the sizes of halos that LIRGs are associated with. However, it is known that the SF properties of galaxy populations in general is also a function of where they are located within their halo environment, for example outskirts vs.\ central regions of clusters \citep{gomesz2003,lewis2002,blanton2007}. Galaxies at the outskirts of galaxy systems tend to have more SF than galaxies falling towards the centre as a result of various processes such as high fly-by interactions \citep{moore1999,sinha2012} and ram-pressure stripping \citep{gunn1972,bower2002} that are effective in shutting off SF. In light of this, to better understand the SF-environment relations, it is beneficial to also identify where the LIRGs dwell inside their respective halos.

We compute the projected distance ($r_\perp$) of each LIRG from the centre of the halo that it is associated with (restricting the sample to halos more massive than $10^{13}M_\odot$), and then normalize by the halo virial radius ($R_{vir}$).  Figure~\ref{fig: dis_cen_includingfield} shows these normalized radii versus the $L_{IR}$ values of the individual galaxies. The different symbols correspond to different halo mass ranges of the galaxy groups. The red diamonds show the mean values of $r_\perp/R_{vir}$ after binning along the $L_{IR}$ axis in a bin size of 0.1 dex. These average points in Figure~\ref{fig: dis_cen_includingfield} reveal that there is a correlation between the projected distance and $L_{IR}$.  Galaxies with high $L_{IR}$, a proxy for the SF rate, are found closer to the centres of their host halos than lower luminosity galaxies.  This average relation is calculated over all halo masses, but breaking it down to smaller halo mass bins (the different symbols) does not reveal any clear trend between the location of LIRGs within their host halo and the host halo mass. 

\section{Discussion and conclusion}

In this work, we have characterized the local environment of LIRG galaxies in terms of the virialized halo that they inhabit, where halos were identified using a galaxy group catalog.  Specifically, we have focused on the masses of the halos that host LIRGs, as well as their locations within these halos.  Based on the mass of the host dark matter halo, the environment of galaxies can be roughly classified into three traditional environmental classes: field, groups, and clusters. Roughly speaking, a dark matter halo with $M_{halo}\leq10^{12.5}M_\odot$ is classified as field; $10^{12.5}M_\odot \leq M_{halo} \leq 10^{13.5}M_\odot$ is classified as a group and $M_{halo}\geq 10^{13.5}M_\odot$ is classified as a cluster \citep{Kauf2004}.  Adopting this classification, we see from the results in Figure \ref{fig: 2mrs_groupmember_bin} that LIRGs preferentially live in groups relative to normal 2MASS galaxies.  Specifically, the fraction of LIRGs in groups is $30-70$\% higher than that of normal galaxies, while the fraction of LIRGs in the field or in clusters is $\sim 50$\% lower than that of normal galaxies.  The significance of this result is limited by the fairly large error bars that are due to the small size of our LIRG sample.  It is important to compile larger volume samples of LIRGs in order to improve on the statistics of this study.

The $L_{IR}$ range of the LIRGs that are found in association with galaxy systems in both target galaxy samples is in a relatively narrow range in between $10^{11}L_\odot$ and $10^{11.6}L_\odot$.  Moreover, the number of galaxies is considerably smaller at the higher $L_{IR}$ end of this range than at the lower end.  Hence, we cannot extend the results for the entire range of LIRGs or into the ULIRG regime.  In previous work using galaxy counts, we established a correlation between the global environmental density and the $L_{IR}$ values spanning both the LIRG and ULIRG regimes \citep{Tekola2012}.  In that work the LIRGs with lower $L_{IR}$ values corresponding to those investigated in this paper turned out to also be in the group regime.  We thus confirm our earlier results by establishing the preferred environment of local LIRGs using an independent technique and, moreover, defining the environment in a physically meaningful way.

While it is clear that ULIRGs tend to be galaxy mergers, or remnants thereof, the reason for the elevated SF of LIRGs in particular has not been well established uniquely since they obviously include apparently isolated galaxies in addition to interacting galaxies, mergers and wide pairs.  Our results both in this work and in \citet{Tekola2012} paint the picture where the larger scale environment of LIRGs is something that sets them apart from normal galaxies.  LIRGs preferentially live in group environments, and they are under-abundant in cluster \textit{and} field environments.  This preferred group size environment supports the hypothesis that star formation in LIRGs is driven by interactions and mergers, as is often observed in individual LIRG cases \citep{sanders1996, hwang2010, ellison2013}, since groups are the environment where most galaxies interact in the present universe.  In the field, the density of galaxies is generally too low for mergers to occur, whereas in massive clusters merger timescales are too long.  An alternative hypothesis perhaps supported by these findings is that star formation in LIRGs is elevated due to larger scale tidal fields of high-density environments \citep{Tekola2012, martig2008}.  

We have also investigated the locations of LIRGs within their host halos and found a trend such that more luminous LIRGs reside closer to the centres of their host halos than galaxies with lower $L_{IR}$.  This is perhaps a further indication that galaxy interactions are responsible for high star formation rates in LIRGs, since interactions are more likely to occur near the halo centre.  We do not however find any correlation of this trend with halo mass, though this may be due to the small size of our LIRG sample.

\end{document}